FULL-LENGTH PAPER

# AlphaFold predicts the most complex protein knot and composite protein knots

Maarten A. Brems[1] | Robert Runkel[1] | Todd O. Yeates[2,3] | Peter Virnau[1]

[1]Department of Physics, Johannes Gutenberg University Mainz, Mainz, Germany

[2]UCLA-DOE Institute for Genomics and Proteomics, University of California Los Angeles, Los Angeles, California, USA

[3]UCLA Department of Chemistry and Biochemistry, University of California Los Angeles, Los Angeles, California, USA

**Correspondence**
Peter Virnau, Department of Physics, Johannes Gutenberg University Mainz, Staudingerweg 9, Mainz 55128, Germany.
Email: virnau@uni-mainz.de

**Funding information**
Deutsche Forschungsgemeinschaft, Grant/Award Number: 233630050-TRR 146; TopDyn; Johannes Gutenberg University Mainz

**Review Editor:** Nir Ben-Tal

**Abstract**

The computer artificial intelligence system AlphaFold has recently predicted previously unknown three-dimensional structures of thousands of proteins. Focusing on the subset with high-confidence scores, we algorithmically analyze these predictions for cases where the protein backbone exhibits rare topological complexity, that is, knotting. Amongst others, we discovered a $7_1$-knot, the most topologically complex knot ever found in a protein, as well several six-crossing composite knots comprised of two methyltransferase or carbonic anhydrase domains, each containing a simple trefoil knot. These deeply embedded composite knots occur evidently by gene duplication and interconnection of knotted dimers. Finally, we report two new five-crossing knots including the first $5_1$-knot. Our list of analyzed structures forms the basis for future experimental studies to confirm these novel-knotted topologies and to explore their complex folding mechanisms.

## 1 | INTRODUCTION

Recently, the artificial intelligence (AI) system AlphaFold developed by Google's DeepMind dominated the Critical Assessment of Techniques for Protein Structure Prediction (CASP) twice.[1] AlphaFold 2, the version under consideration here, is a deep learning system that incorporates training procedures based on the evolutionary, physical, and geometric constraints of protein structures.[2,3] It features iterative refinement of predictions and allows for learning from unlabeled protein sequences using self-distillation and self-estimates of accuracy to directly predict the 3D coordinates of all heavy atoms for a given protein using the primary structure and aligned sequences of homologues.[2,3] AlphaFold 2 has currently predicted several hundred thousand protein structures, most of which are not contained in the Protein Data Bank (PDB),[4,5] which mainly archives experimentally determined structures.[5] Thereby, AlphaFold's prediction databank may be of tremendous value, especially for the research of protein phenomena which are infrequent but still of high relevance to understand the intricacies of the underlying mechanisms of protein folding.

A particularly fascinating phenomenon arises for proteins that contain a topological knot in their polypeptide backbone,[6–30] that is, proteins which would not fully disentangle after being pulled from both ends.[6] In the past two decades, only about 20 different protein families

---

Maarten A. Brems and Robert Runkel have contributed equally.

containing knots have been identified.[7] Nevertheless, knotted proteins pose a challenge to protein folding and evolution.[8] Simulation algorithms often overestimate the knotting probability of proteins as the latter is lower than the knotting probability of random chains.[8,12,14,31] Moreover, protein topology is usually similar among homologues, meaning that knotted folds tend to be preserved across proteins closely related in evolution. For these reasons, and owing to the established rarity of knotting among natural proteins, the potential presence of knotted topologies in the vast new database of predicted protein structures is of keen interest. Currently, the most complex knot found in a protein is a single knot with six essential crossings in any projection to a plane[32]; a composite knot has not been observed yet.

We searched the entire AlphaFold 2 databank, including the "Model organism proteomes", "Swiss-Prot" and "Global health proteomes" data sets,[4] for topologically complex proteins containing previously unknown deep knots (which still persist when at least five amino acids [aa] are cut from both ends). We excluded from the analysis those with lower confidence scores (<80) or exceedingly long protein chains (>600 aa), where predicted accuracy and ability to experimentally validate the structures could be limiting. The applied criteria for the survey as well as our knot detection algorithm are discussed in detail in the methods section. Based on this search and visual inspection, we have identified the first $7_1$-knot (with at least seven crossings in any projection onto a plane) as well as a likely evolutionary mechanism for generation of $3_1\#3_1$ composite knots, accompanied by several examples. Moreover, we report two new five-crossing knots including the first $5_1$-knot in a protein and provide an overview of additional knotted proteins present in the AlphaFold databank (Supporting Information S1).

## 2 | GENERATION MECHANISM OF COMPOSITE KNOTS

Our survey identified nine cases of composite knots, previously unknown. These are all instances where two essentially independent trefoil knots are present in one longer protein chain. We propose a novel mechanism for generation of such composite knots based on gene duplication and the interconnection of a knotted homodimer. Interestingly, this mechanism resembles a strategy employed for the creation of the first artificial protein knot in which an unknotted dimer was "connected" to form a trefoil.[17] We have observed multiple instances including the methyltransferases and carbonic anhydrases, as discussed below, in which proteins containing

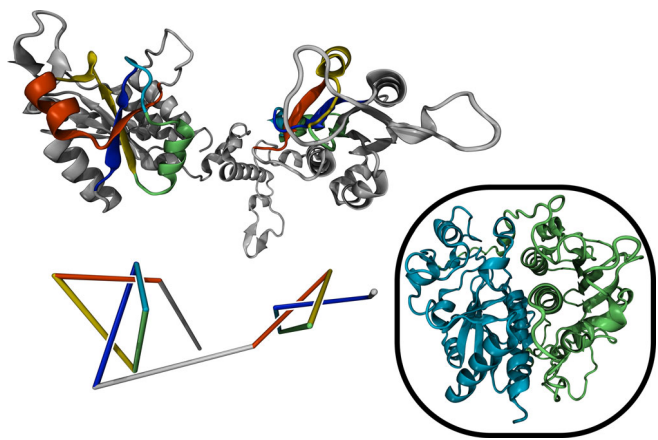

**FIGURE 1** 3D structure (top) and reduced representation (bottom) of a six-crossing composite knot in protein Q313J9 (methyltransferase). A composite trefoil knot ($3_1\#3_1$) can be identified. Topologically trivial segments are not displayed. Inset: A similar structure is predicted for Methyltransferase A4I142, except the two knotted domains form a more compact arrangement

a composite trefoil knot ($3_1\#3_1$) are homologous to a known knotted homodimer with one trefoil knot in each chain. Figure 1 depicts protein Q313J9, which has been identified as tRNA (guanine-N1-)-methyltransferase, with a length of 425 aa and a knotted core between residues 86 and 360. If not stated otherwise, the protein code refers to the UniProt/AlphaFold identifier[33] and structures are visualized using the Visual Molecular Dynamics (VMD) software.[34] To visualize knots in the protein structures, we employ reduced representations (bottom structures in Figures 1-4), in which the protein is divided into segments such that topology is conserved when the segments are replaced by straight lines connecting their respective start and end points. Methyltransferases are known to usually contain a single trefoil knot[7] per chain and sometimes appear as homodimers. We have observed two variations of this phenomenon: For protein Q313J9 in Figure 1 and a similar methyltransferase Q72DU3, the two main segments containing the trefoil knots appear flexibly connected, whereas predictions for some proteins of similar sequence preserve the presumed original dimer structure more strictly. (See inset of Figure 1.) Examples are the methylases A4I142, Q4DMW6, and Q4D5S2 as well as proteins Q4CYG6, Q4D7N4, and Q381U1. The latter are labeled as uncharacterized but show about 15% sequence identity and 30% matching secondary structure with the methyltransferase pdb:2ha8:A for proteins Q4CYG6 and Q4D7N4 or with the methyltransferase pdb:1v2x:A for protein Q381U1 according to the PBDeFold webserver.[35] Structural alignment and sequence identity discussions based on PDBeFold[35,36] for each group of methyltransferases containing composite knots can be found in the Supporting Information (SI). A particularly interesting example of the second variant is Carbonic anhydrase P54212 (Figure 2), with a length of 589 aa and a knotted core between residues 198 and 570. Carbonic anhydrases were the first proteins identified as being knotted.[37] Both trefoil knots in methyltransferase Q313J9 and as well as in carbonic anhydrase P54212 have positive chirality. Therefore, the composite trefoil knots can be identified as what is commonly known as a granny knot.[38] The chirality of the composite knots is in agreement with previous results reporting positive chirality for the single trefoil knots in methyltransferases and carbonic anhydrases.[7] We have thereby observed the same phenomenon, a potential mechanism for generation of composite knots, in two distinct protein families and with two structural variations.

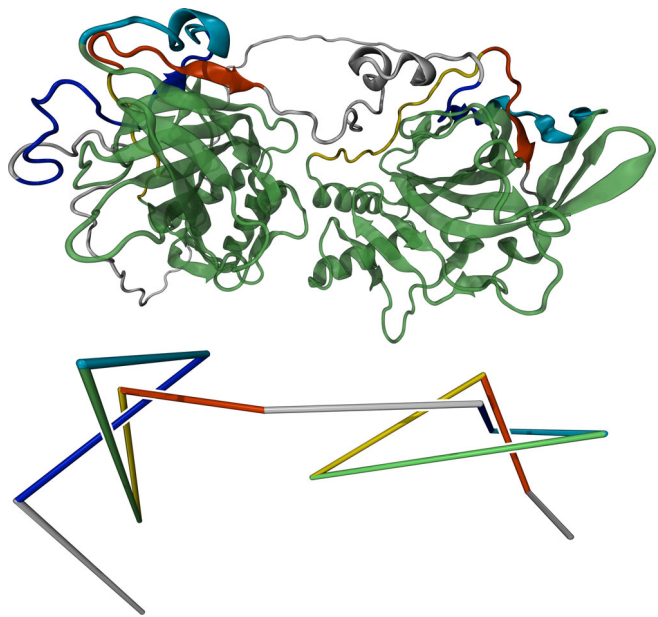

**FIGURE 2** 3D structure (top) and reduced representation (bottom) of protein P54212 (carbonic anhydrase). A composite trefoil knot ($3_1\#3_1$) can be identified. Topologically trivial segments are not displayed. The large green segments in the top structure are made transparent for a better view of the knotted region

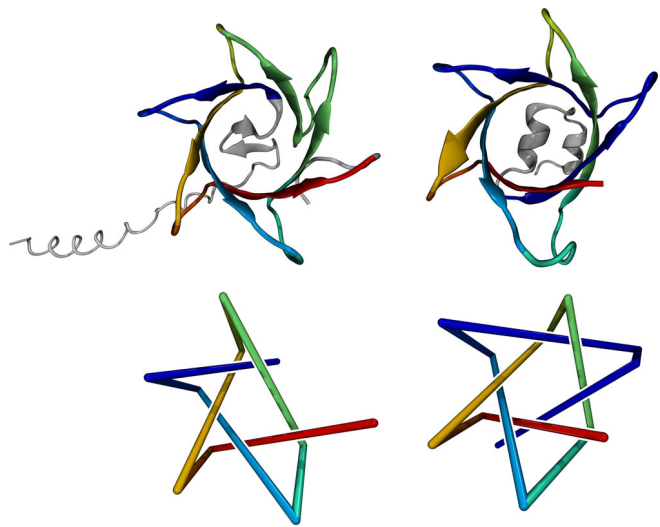

**FIGURE 3** Structure and topology of proteins P73136 (left) and Q9PR55 (right). Top: 3D structures predicted by AlphaFold. Bottom: Reduced representations to visualize the $5_1$- and $7_1$-knots in the left and right structure, respectively. On the right, the dark blue segment introduces an additional winding

## 3 | FIRST $7_1$-KNOT IN A PROTEIN

Figure 3 depicts proteins P73136 and Q9PR55 with lengths of 112 and 89 amino acids, respectively. Both are uncharacterized and no probable homologues could be identified using PDBeFold. However, they have 48% sequence identity and 71% matching secondary structure with respect to each other, which indicates that they are probably homologues. Protein Q9PR55 contains the most complicated knot, a $7_1$-knot, known to date with a knotted core between residues 27 and 83. The similar structure of protein P73136 contains a $5_1$-knot with a knotted core between residues 45 and 94. Such a pair of homologues where the two proteins possess a different nontrivial topology has not been observed previously. A closer look reveals that the more complex topology of protein Q9PR55 arises from a protein segment that introduces an additional winding (dark blue in Figure 3, right); a $7_1$-torus knot is essentially a $5_1$-torus knot with one additional winding around the torus. Both knots have positive chirality.

## 4 | NEW $5_1$- AND $5_2$-KNOTS

We have found two previously unknown knots with five essential crossings, including the first $5_1$-knot. Figure 4 (left) depicts protein A0A0K0IQS9 (Bm1115) which contains a $5_1$-knot. Its length is 173 aa and its knotted core extends from residue 39 to residue 157. Protein C1GYM9 (Figure 4 right) is uncharacterized, and no probable homologue could be identified using PDBeFold. It contains a $5_2$-knot with a knotted core between residues 76 and 391 and its length is 420 aa. Both knots exhibit positive chirality.

## 5 | TESTS OF ACCURACY

Owing to the novelty of the findings here, validation by independent methods will be important. Ahead of experimental studies, here we applied an orthogonal computational tool, ERRAT,[39] to assess the predicted knotted

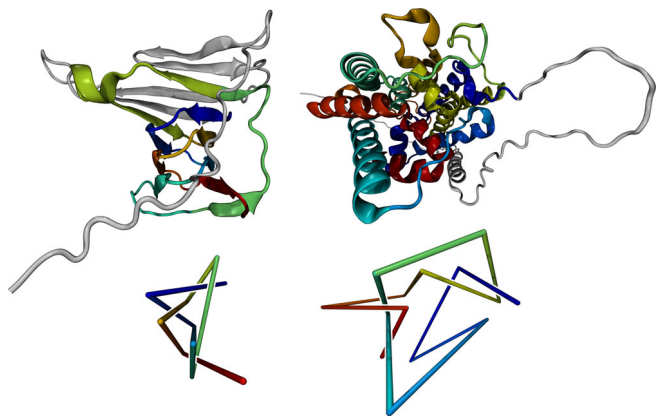

**FIGURE 4** Structure and topology of proteins A0A0K0IQS9 (left) and C1GYM9 (right). Top: 3D structures predicted by AlphaFold. Bottom: Reduced representations to visualize the $5_1$- and $5_2$-knots in the left and right structure, respectively

structures. The ERRAT algorithm evaluates patterns of non-bonded contacts between C, N, and O atoms, and makes a statistical comparison to high resolution structures. By being distinct from metrics employed in AlphaFold (and other prediction methods), it offers an independent assessment. We ran ERRAT[40] on the set of knotted structures discussed above. Discounting occasional extended termini found in some models, all the models tested showed good scores; all cases have >90% of their protein chain falling within (below) the 95% threshold for rejecting unlikely conformations. Our overall assessment was therefore that the predicted structures are correct, at least to a large extent. However, in some cases, local regions of structure appeared potentially problematic. And it is critical to note that minor discrepancies in the path of a protein chain—for example, those that would change an over/under crossing—can change the topology, potentially leading to an incorrect assignment of a knot. With regard to the present study, we note that, for the composite knot Q4D5S2 (and its relatives), the ERRAT program flags a beta strand segment around residues 100–110 as likely to be structurally incorrect (SI, Figure S1). Notably, the passage of the chain in this region is important for the knotted topology. While the AlphaFold program assigns a high degree of confidence to the predicted structure in this region, our independent assessment emphasizes the need for confirmatory experimental studies.

## 6 | DISCUSSION AND CONCLUSIONS

In conclusion, we have analyzed all predictions for protein 3D structures by the AlphaFold AI system for new topologically complex proteins. Our complete analysis of the data provided by AlphaFold (see SI) reveals several high-confidence proteins containing deep complex knots, which are suitable for experimental verification of their 3D structure. In this data set, we found amongst others a $7_1$-knot, the most complex ever discovered in a protein, as well as a new $5_1$-knot in a homologue structure and the first instances of composite protein knots. For the latter, we propose an evolutionary mechanism for their creation by gene duplication. As protein topology is an ongoing challenge for protein folding algorithms, it will be important to verify or refute the discussed structure predictions experimentally. One would not only obtain a fine gauge for the capability of AlphaFold AI system to correctly predict the topology of complex proteins, but importantly confirm the multitude of novel protein knots identified here.

## 7 | METHODS

Mathematically, knots are well-defined in closed three-dimensional curves, and can be categorized according to the minimal number of crossings the curve makes in a projection onto a plane, allowing for any non-breaking manipulations (e.g., smoothing) of the curve. The simplest non-trivial knot is the so-called trefoil knot with three crossings. The figure-eight knot has four crossings, there are two knots with five, three knots with six, and eight distinct knots with seven crossings. In addition, simpler knots can be combined—i.e., formed on separable regions of the same curve—to form composite knots, which are distinct from prime knots; the latter cannot be decomposed into simpler knots. In the present study, topologically non-trivial proteins (i.e., polypeptide backbones that are knotted) have been identified using a classification algorithm based on the Alexander polynomial invariant.[41,42] Note that for a knot to be well-defined, the two ends of the protein must be virtually closed,[41,42] which sometimes leads to ambiguous results and requires additional visual inspection. Employing the algorithm above, we find that the knotting probability (of around 2%) of the AlphaFold database is roughly in accordance with the one from PDB as discussed in the SI.

We limit our detailed, non-algorithmic analysis to proteins which fulfill the following three criteria: First, the average computed confidence score for the predicted structure must be 80 or above. The AlphaFold AI system provides a per-residue estimate of its confidence on a scale from 0 to 100, which is based on the lDDT-Cα metric.[43] Second, the topology of the protein must be more complex than a trefoil ($3_1$) and figure-eight ($4_1$) knot, that is, it must contain a knot with at least five essential crossings, which includes any potential composite knots. We exclude combinations of knot types and protein families which are already known, such as $5_2$-knots in ubiquitin hydrolases and $6_1$-knots in haloacid dehalogenase.[7] Moreover, the knot must be deep in the sense that the topology of the system is invariant under removal of at least 5 aa from both termini. A related measure for the topological robustness of a structure, which we employ in our discussions, is the extend of the knotted core, that is, the smallest region of the protein which still contains the knot. The extend of the knotted core is one of the measures included in the knot matrix representation for proteins introduced by King et al. in Ref. [18] and popularized in further work.[44] Third, the protein must not exceed a length of 600 aa. The final condition was set to mitigate the potential errors in topology assignment that can arise from relatively small structural discrepancies in large structures, in addition to challenges typically associated with experimental studies on very large and potentially flexible protein chains. As established above, correct prediction of protein topology is still an important challenge for modern computer algorithms. Thus, ultimate experimental verification or refutation will highlight the degree to which the AlphaFold AI system can grasp the intricacies of protein folding for highly complex cases. An extensive table of all knotted proteins in AlphaFold's databank, as determined by our algorithm, including all quantitative measures employed in our analysis and filtering can be found in the SI. In the present work, the most interesting proteins that fulfill the conditions above are discussed in detail. In the SI, we also list proteins that fulfilled the computational criteria, but which were set aside as potentially unreliable after visual inspection. The per-residue confidence scores of all proteins depicted in the figures are given in the SI; we observe that no segments which are substantial for the knots possess particularly low confidence. Moreover, we want to acknowledge that we found the $6_3$-knot in von Willebrand factor A (identifiers O00534 and Q99KC8), which was also reported in Ref. [45], where AlphaFold predictions for the human proteome were studied, even though it does not satisfy the above conditions stated above due to its length. In the review stage of this manuscript, another paper was published by the same group, which describes a server to determine knots in predicted structures from AlphaFold.[46]

## AUTHOR CONTRIBUTIONS

**Maarten Alexander Brems:** Formal analysis (equal); investigation (supporting); software (equal); visualization (lead); writing – original draft (lead); writing – review and editing (equal). **Robert Runkel:** Formal analysis (equal); investigation (lead); software (equal); writing – review and editing (supporting). **Todd Yeates:** Conceptualization (supporting); methodology (supporting); project administration (supporting); software (equal); supervision (supporting); visualization (supporting); writing – original draft (supporting); writing – review and editing (supporting). **Peter Virnau:** Conceptualization (lead); funding acquisition (lead); methodology (lead); project administration (lead); resources (lead); supervision (lead); writing – review and editing (equal).

## ACKNOWLEDGMENTS

We are grateful to the Deutsche Forschungsgemeinschaft (DFG, German Research Foundation) for funding this research: Project number 233630050-TRR 146. The authors gratefully acknowledge computing time granted on the HPC cluster Mogon at Johannes Gutenberg University Mainz. The authors furthermore acknowledge funding from TopDyn. Open Access funding enabled and organized by Projekt DEAL.

## CONFLICT OF INTEREST

There is no conflict of interest to declare.

## ORCID

*Maarten A. Brems*  https://orcid.org/0000-0002-0210-508X
*Todd O. Yeates*  https://orcid.org/0000-0001-5709-9839
*Peter Virnau*  https://orcid.org/0000-0003-2340-3884

## REFERENCES

1. Callaway E. 'It will change everything': DeepMind's AI makes gigantic leap in solving protein structures. Nature. 2020;588: 203–204.
2. Jumper J, Evans R, Pritzel A, et al. Highly accurate protein structure prediction with AlphaFold. Nature. 2021;596:583–589.
3. Varadi M, Anyango S, Deshpande M, et al. AlphaFold protein structure database: Massively expanding the structural coverage of protein-sequence space with high-accuracy models. Nucleic Acids Res. 2022;50:D439–D444.
4. Anon AlphaFold Protein Structure Database. Available from: https://alphafold.ebi.ac.uk/
5. wwPDB consortium. Protein data Bank: The single global archive for 3D macromolecular structure data. Nucleic Acids Res. 2019;47:D520–D528.
6. Virnau P, Mallam A, Jackson S. Structures and folding pathways of topologically knotted proteins. J Phys Condens Matter. 2011;23:033101.
7. Jarmolinska AI, Perlinska AP, Runkel R, et al. Proteins' knotty problems. J Mol Biol. 2019;431:244–257.
8. Virnau P, Mirny LA, Kardar M. Intricate knots in proteins: Function and evolution. PLoS Comput Biol. 2006;2:e122.
9. Mansfield ML. Are there knots in proteins? Nat Struct Mol Biol. 1994;1:213–214.
10. Mansfield ML. Fit to be tied. Nat Struct Mol Biol. 1997;4:166–167.
11. Taylor WR. A deeply knotted protein structure and how it might fold. Nature. 2000;406:916–919.

12. Lua RC, Grosberg AY. Statistics of knots, geometry of conformations, and evolution of proteins. PLoS Comput Biol. 2006; 2:e45.
13. Yeates TO, Norcross TS, King NP. Knotted and topologically complex proteins as models for studying folding and stability. Curr Opin Chem Biol. 2007;11:595–603.
14. Potestio R, Micheletti C, Orland H. Knotted vs. unknotted proteins: Evidence of knot-promoting loops. PLoS Comput Biol. 2010;6:e1000864.
15. Taylor WR, Lin K. Protein knots: A tangled problem. Nature. 2003;421:25.
16. Kamitori S. A real knot in protein. J Am Chem Soc. 1996;118: 8945–8946.
17. King NP, Jacobitz AW, Sawaya MR, Goldschmidt L, Yeates TO. Structure and folding of a designed knotted protein. Proc Natl Acad Sci. 2010;107:20732–20737.
18. King NP, Yeates EO, Yeates TO. Identification of rare slipknots in proteins and their implications for stability and folding. J Mol Biol. 2007;373:153–166.
19. Sriramoju MK, Chen Y, Lee Y-TC, Hsu S-TD. Topologically knotted deubiquitinases exhibit unprecedented mechanostability to withstand the proteolysis by an AAA+ protease. Sci Rep. 2018;8:7076.
20. Ko K-T, Hu I-C, Huang K-F, Lyu P-C, Hsu S-TD. Untying a knotted SPOUT RNA methyltransferase by circular permutation results in a domain-swapped dimer. Structure. 2019;27: 1224–1233.e4.
21. Jamroz M, Niemyska W, Rawdon EJ, et al. KnotProt: A database of proteins with knots and slipknots. Nucleic Acids Res. 2015;43:D306–D314.
22. Sulkowska JI, Sulkowski P, Onuchic J. Dodging the crisis of folding proteins with knots. Proc Natl Acad Sci USA. 2009;106: 3119–3124.
23. Sulkowska JI, Sulkowski P, Szymczak P, Cieplak M. Stabilizing effect of knots on proteins. Proc Natl Acad Sci USA. 2008;105: 19714–19719.
24. Faisca PFN. Knotted proteins: A tangled tale of structural biology. Comp Struct Biotechnol J. 2015;13:459–468.
25. Faisca PFN, Travasso RDM, Charters T, Nunes A, Cieplak M. The folding of knotted proteins: Insights from lattice simulations. Phys Biol. 2010;7:016009.
26. Jackson SE, Suma A, Micheletti C. How to fold intricately: Using theory and experiments to unravel the properties of knotted proteins. Curr Opin Struct Biol. 2017;42:6–14.
27. Mallam AL, Jackson SE. Folding studies on a knotted protein. J Mol Biol. 2005;346:1409–1421.
28. Mallam AL, Jackson SE. Knot formation in newly translated proteins is spontaneous and accelerated by chaperonins. Nat Chem Biol. 2012;8:147–153.
29. Mallam AL, Jackson SE. Probing Nature's knots: The folding pathway of a knotted homodimeric protein. J Mol Biol. 2006; 359:1420–1436.
30. Lim NCH, Jackson SE. Molecular knots in biology and chemistry. J. Phys.: Condes. Matter. 2015;27:354101.
31. Wüst T, Reith D, Virnau P. Sequence determines degree of Knottedness in a coarse-grained protein model. Phys Rev Lett. 2015;114:028102.
32. Bölinger D, Sułkowska JI, Hsu H-P, et al. A Stevedore's protein knot. PLoS Comput Biol. 2010;6:e1000731.
33. The UniProt Consortium. UniProt: The universal protein knowledgebase in 2021. Nucleic Acids Res. 2021;49:D480–D489.
34. Humphrey W, Dalke A, Schulten K. VMD: Visual molecular dynamics. J Mol Graph. 1996;14(33–38):27–28.
35. Krissinel E, Henrick K. Secondary-structure matching (SSM), a new tool for fast protein structure alignment in three dimensions. Acta Cryst D. 2004;60:2256–2268.
36. Krissinel E, Henrick K. Multiple alignment of protein structures in three dimensions. In: Berthold MR, Glen RC, Diederichs K, Kohlbacher O, Fischer I, editors. Computational life sciences, Lecture notes in computer science. Berlin, Heidelberg: Springer, 2005; p. 67–78.
37. Richardson JS. β-Sheet topology and the relatedness of proteins. Nature. 1977;268:495–500.
38. Rolfsen D. Knots and links. Berkeley, CA: Publish or Perish, 1976.
39. Colovos C, Yeates TO. Verification of protein structures: Patterns of nonbonded atomic interactions. Protein Sci. 1993;2: 1511–1519.
40. Anon SAVESv6.0—Structure Validation Server. Available from: https://saves.mbi.ucla.edu/
41. Virnau P. Detection and visualization of physical knots in macromolecules. Phys Procedia. 2010;6:117–125.
42. Kolesov G, Virnau P, Kardar M, Mirny LA. Protein knot server: Detection of knots in protein structures. Nucleic Acids Res. 2007;35:W425–W428.
43. Mariani V, Biasini M, Barbato A, Schwede T. lDDT: A local superposition-free score for comparing protein structures and models using distance difference tests. Bioinformatics. 2013;29: 2722–2728.
44. Sułkowska JI, Rawdon EJ, Millett KC, Onuchic JN, Stasiak A. Conservation of complex knotting and slipknotting patterns in proteins. Proc Natl Acad Sci USA. 2012;109:E1715–E1723.
45. Perlinska AP, Niemyska WH, Gren BA, Rubach P, Sulkowska JI (2022) New 63 knot and other knots in human proteome from AlphaFold predictions: December 30, 2021.474018. Available from: https://doi.org/10.1101/2021.12.30.474018v1
46. Niemyska W, Rubach P, Gren BA, et al. AlphaKnot: Server to analyze entanglement in structures predicted by AlphaFold methods. Nucleic Acids Res.: gkac388. 2022.

## SUPPORTING INFORMATION

Additional supporting information can be found online in the Supporting Information section at the end of this article.